\documentclass{PoS}

\usepackage{amsfonts}
\usepackage{amsmath}
\usepackage{subfigure} 

\newcommand{\TB}{\textrm{B}}
\newcommand{\TI}{\textrm{I}}
\newcommand{\TP}{\textrm{P}}
\newcommand{\TV}{\textrm{V}}
\newcommand{\TA}{\textrm{A}}
\newcommand{\TT}{\textrm{T}}

\title{
  Determination of $B_K$ using improved staggered fermions
  (II) SU(2) chiral perturbation theory fit
}

\ShortTitle{
  Determination of $B_K$ using improved staggered fermions (II) SU(2) fit
}

\author{\speaker{Hyung-Jin Kim}, Taegil Bae, Jangho Kim,
  Jongjeong Kim, Kwangwoo Kim, Boram Yoon, Weonjong Lee\\
  Frontier Physics Research Division and Center for Theoretical Physics \\
  Department of Physics and Astronomy, 
  Seoul National University, Seoul, 151-747, South Korea \\
  E-mail: \email{wlee@snu.ac.kr}}

\author{Chulwoo Jung \\
  Physics Department, Brookhaven National Laboratory,
  Upton, NY11973, USA \\
  E-mail: \email{chulwoo@bnl.gov}}

\author{Stephen R. Sharpe\\
  Physics Department, University of Washington, Seattle, WA 98195-1560 \\
  E-mail: \email{sharpe@phys.washington.edu}}

\abstract{We present results for $B_K$ calculated using HYP-smeared 
  improved staggered fermions on the MILC asqtad lattices.
  In this report, the data is analyzed using the results of
  SU(2) staggered chiral perturbation theory (SChPT).
  We outline the derivation of the NLO SU(2) SChPT result,
  explain our fitting procedure, and outline how we estimate
  systematic errors. We also show the light sea-quark mass and lattice
  spacing dependence for both SU(2) and SU(3)-based analyses.
  Our preliminary result from the SU(2) analysis is 
  $B_K(\text{NDR}, \mu   = 2 \text{ GeV}) = 0.512 \pm 0.014 \pm 0.034$
  and $\hat{B}_K = B_K(\text{RGI})= 0.701 \pm 0.019 \pm 0.047$. 
  This is somewhat more accurate than our result from the SU(3) analysis.
  It is consistent with results obtained using valence domain-wall fermions.}

\FullConference{
  The XXVII International Symposium on Lattice Field Theory - LAT2009\\
  July 26-31 2009
  Peking University, Beijing, China
}

\begin{document}

\section{Introduction} 
This paper is the second in a series of the four proceedings 
describing our calculation of $B_K$.
In the first, we provided a brief phenomenological introduction,
outlined the results of SU(3) staggered chiral perturbation theory (SChPT),
and explained the corresponding fitting strategy~\cite{ref:wlee:2009-1}.
Here, we explain briefly how we obtain the SU(2) SChPT result,
outline how we use this result to analyze our data, and 
present a preliminary value, and error budget, for $B_K$.
The details of the lattice ensembles
and quark masses are as in Ref.~\cite{ref:wlee:2009-1}.

The use of SU(2) ChPT was pioneered in the lattice context
by the RBC collaboration~\cite{ref:RBCSU2}. 
One treats kaons and $\eta$'s
as heavy, static sources for pseudo-Goldstone bosons (PGBs) 
composed of light quarks (the ``pions''). Unlike the SU(3) version,
SU(2) ChPT does not require an expansion in the
ratio $r_s=m_s/\Lambda_{\rm QCD}$.
The expansion parameter is thus
$r_l = m_\ell/\Lambda_{\rm QCD}$ (with $m_\ell$ the common
up and down quark mass in our isospin-symmetric simulations).
This improves convergence properties (as long as $m_\ell$
is light enough), although this comes at a price:
one fits to a smaller number of data-points, 
and must do so with low-energy coefficients (LECs)
that have an unknown, analytic dependence on $m_s/\Lambda_{\rm QCD}$.

Our calculations have valence quark masses ranging from approximately
$m_s^{\rm phys}/10$ to $m_s^{\rm phys}$.
Despite the large upper value, we find that SU(3) fits give a 
reasonable description of our data. The main problem is the
presence of multiple contributions to the NLO formula
arising from discretization errors or from truncation of the 
perturbative matching factors. 
These lattice artefacts are poorly determined in the fits
and lead to $\sim 5\%$ uncertainties upon extrapolation to the
physical light-quark mass.
A major advantage of the SU(2) approach in our context is that
all these artefacts are moved to NNLO, i.e. are known to be very
small. This allows a better-controlled extrapolation to the physical
mass.

\section{SU(2) Staggered ChPT Analysis}
The following description is necessarily very brief. A full
explanation will appear in Ref.~\cite{ref:future}.

We need to generalize the SU(2) result for $B_K$ given
in continuum ChPT in Ref.~\cite{ref:RBCSU2} to include the
artefacts associated with staggered fermions. This generalization
has been done in the SU(3) case in Ref.~\cite{ref:sharpe:1},
but requires a rather involved operator enumeration.\footnote{%
Our use of a mixed action leads to small corrections to the
analysis of Ref.~\cite{ref:sharpe:1}
that we have determined~\cite{Baelat08,ref:wlee:2009-1,ref:future}.}
Rather than carry out a direct generalization to SU(2) SChPT, we instead
have worked to all orders in $r_s$ (though to NLO in $r_\ell$)
within SU(3) SChPT. 
Using power-counting arguments,
we find a simple result~\cite{ref:future}: 
one can obtain the NLO SU(2) SChPT result simply by taking
the SU(2) limit of the NLO SU(3) SChPT result, 
and then  allowing 
(almost) all of the LECs to have an unknown dependence on $r_s$.
The only exception to this arbitrary
dependence is the factor of $1/f_\pi^2$ multiplying the pion chiral
logarithms, which remains unchanged.

Applying this result we find the simplification noted above,
namely that {\em all} the non-analytic contributions multiplied
by LECs proportional
to $a^2$ or $\alpha^2$ are pushed to NNLO in SU(2) SChPT. This happens
because the factor of $M_K^2$ in the numerator of $B_K$ is 
balanced by a chiral logarithm proportional to $M_\pi^2 \log(M_\pi)$.
In SU(3) ChPT this ratio is of $O(1)$, but in the SU(2) case it
is of NLO. The overall factor of $a^2$ or $\alpha^2$ then moves
the term to NNLO.

The final result is that taste-breaking effects only enter 
the NLO expression through the masses of the pions of different tastes.
The predicted form is
\begin{eqnarray}
  f_\text{th} &=& d_0 F_0 + d_1 \frac{X_P}{\Lambda_\chi^2}  + d_2
  \frac{X_P^2}{\Lambda_\chi^4} + d_3 \frac{L_P}{\Lambda_\chi^2}
\label{eq:fth}
\end{eqnarray}
where the chiral logarithms (defined in Ref.~\cite{ref:wlee:2009-1})
reside in the function
\begin{eqnarray}
  F_0 &=& 1 + \frac{1}{32\pi^2 f_\pi^2} \Big\{
  \ell(X_\TI) + (L_\TI-X_\TI)\tilde\ell(X_\TI)
  - 2 \langle \ell(X_\TB) \rangle \Big\}
  \\
  \langle \ell(X_\TB)\rangle &=&
  \frac{1}{16} \Big[
    \ell(X_\TI) + \ell(X_\TP) + 4 \ell(X_\TV)
    +4 \ell(X_\TA) + 6 \ell(X_\TT) \Big]\,.
\end{eqnarray}
Here $X_\TB$ ($L_\TB$) is the squared mass
of the valence (sea) pion with taste B, which we know
from our simulations or those of the MILC collaboration.
The scale $\Lambda_\chi$ is arbitrary and we take it to be 1 GeV.
The coefficients $d_i$ have an unknown dependence 
on $r_s$, and, in addition, at NLO $d_0$ depends also on $a^2$
and $\alpha^2$.

As for SU(3) fitting, we include a single analytic NNLO 
term---that proportional to $d_2$. 
In the SU(2) case, however, we find that we can drop this term if
we consider only the smallest valence light-quark masses.
In the following we label such fits as ``NLO'', while if we
include the $d_2$ term we label the fits with ``NNLO''.

\section{SU(2) SChPT Fitting}

A major advantage of the SU(2) analysis is that the fitting function
is much simpler than that for SU(3).
As a consequence, we do not need to include Bayesian priors
for any of the parameters.

Our choice of which valence quark masses to include in the
analysis is exemplified by our approach on the coarse MILC lattices
($a\approx 0.12\;$fm).
Here the physical strange quark mass is $a m^\textrm{phy}_s \cong
0.052$, and we choose the nearest three values
for the valence strange-quark mass, 
$a m_y = \{0.05, 0.045, 0.04\}$, in order to extrapolate to the
physical value. For the valence down-quark mass we use our
lightest four values, 
$am_x = \{0.005, 0.01, 0.015, 0.02\}$, to extrapolate to $am_d^{\rm phys}$.
These choices ensure that the expansion parameter of
SU(2) ChPT is relatively small: $m_x/m_y \le 1/2$.
Analogous choices of quark masses are made on the fine and
superfine ensembles~\cite{ref:future}.
We call this choice ``4X3Y'', and use it for our central value.
We have also considered ``5X3Y'' fits.

\begin{figure}[t!]
\centering
    \includegraphics[width=0.49\textwidth]
                    {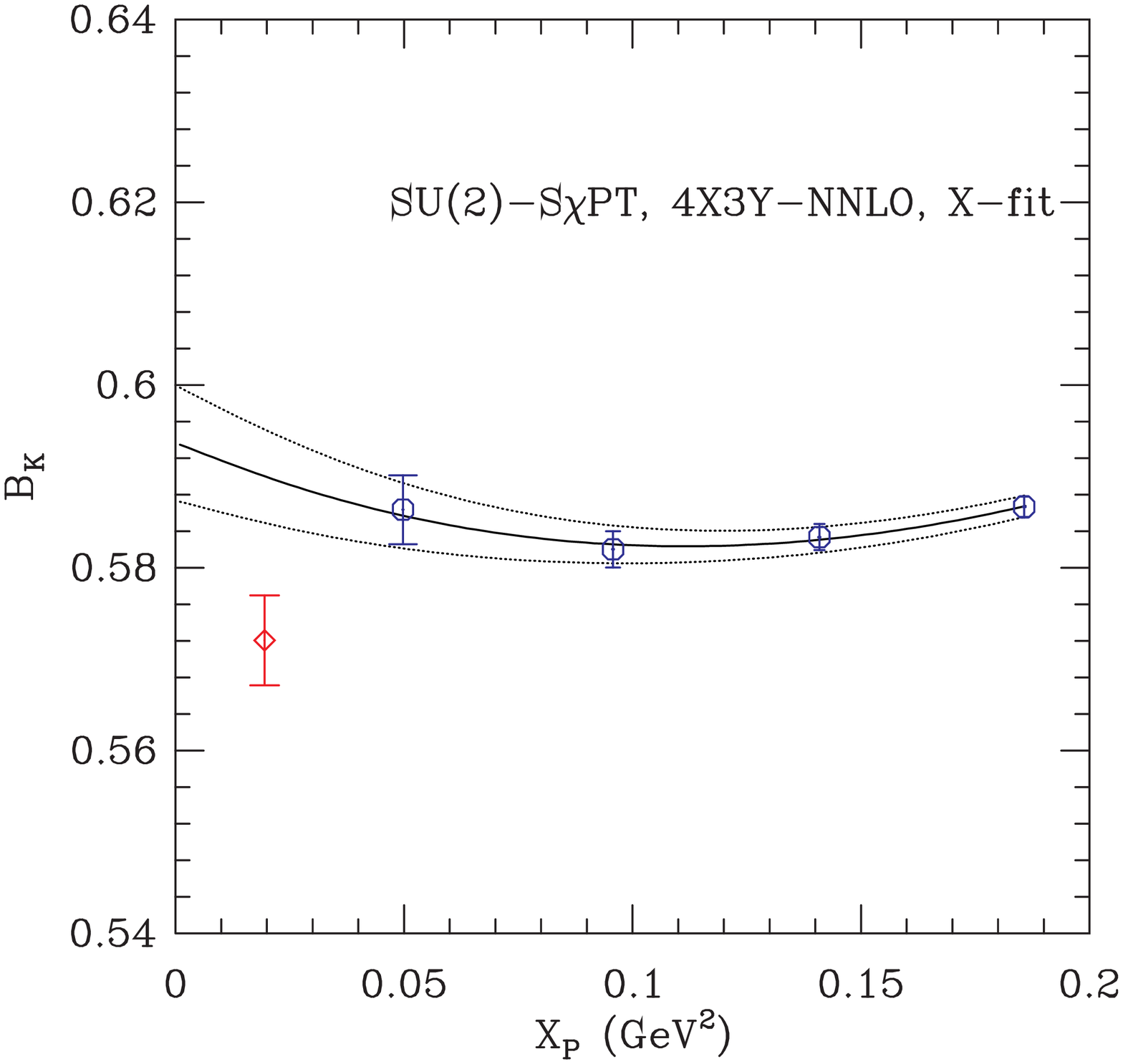}
    \includegraphics[width=0.49\textwidth]
                    {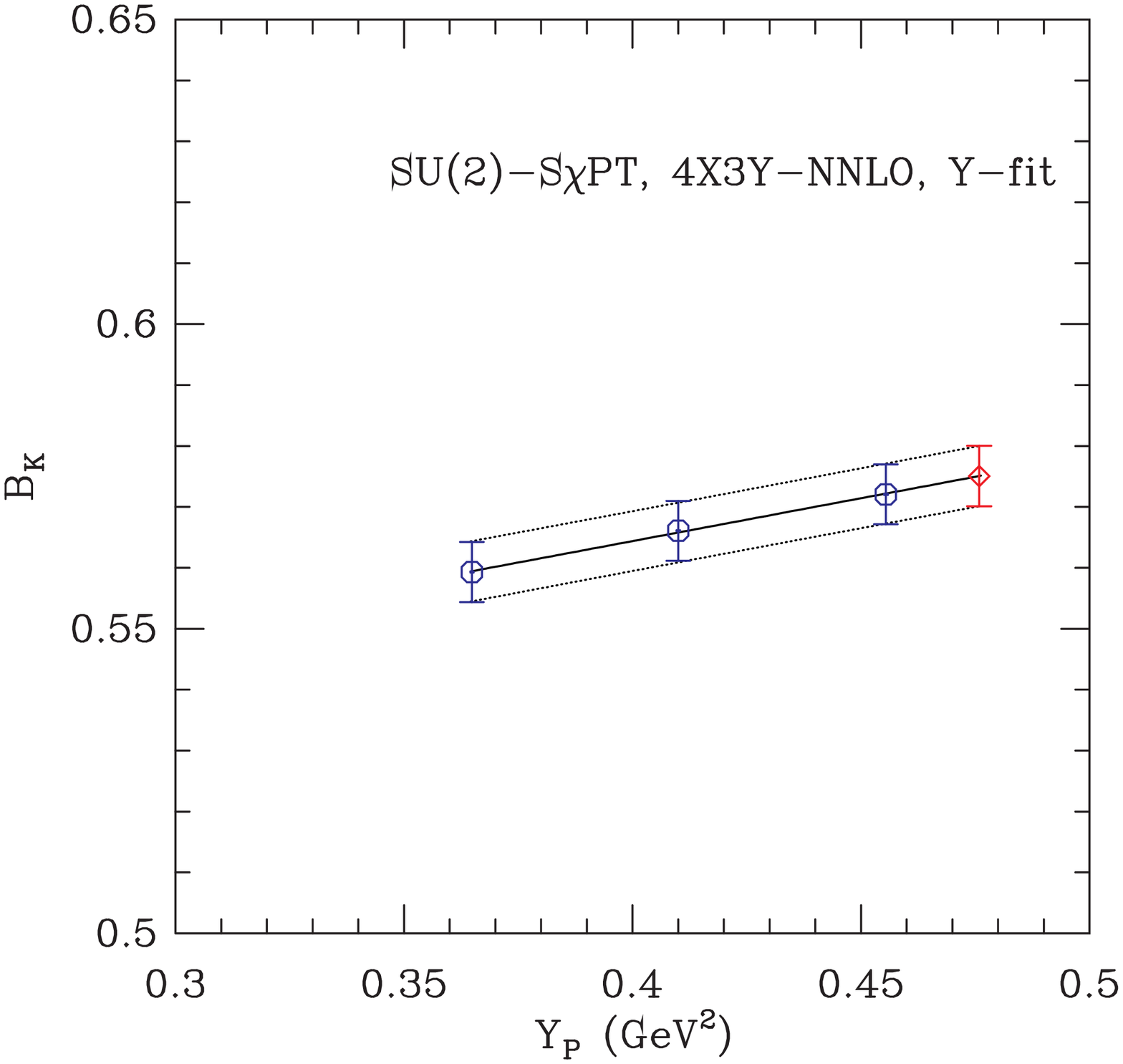}
\caption{One-loop matched $B_K$ fitted versus $X_P$ (left) and $Y_P$ (right), 
  on the MILC coarse lattices with $am_\ell=0.01$ and
  $am_s=0.05$. In the left panel $am_y=0.05$, and the (red) diamond
  shows the result after removing lattice artefacts.
  The fit type is 4X3Y-NNLO. $B_K$ is
  obtained using one-loop matching. See text for more details.}
\label{fig:su2-4x3y-nnlo}
\end{figure}
In Fig.~\ref{fig:su2-4x3y-nnlo}, we show examples of the resulting fits.
In the left plot, the (blue) octagons show the one-loop matched
lattice data, which are then fit to the NNLO form of
eq.~(\ref{eq:fth}). There are only 3 parameters because
$L_P$ is fixed in this partially-quenched fit.
With the fit parameters determined, we then evaluate the
expression (\ref{eq:fth}) at the physical
pion mass {\em and with taste splittings set to zero},
i.e with $X_\TB=L_\TI=M_\pi^2$.
This removes all taste-breaking discretization and truncation errors,
and results in the point shown with the (red) diamond.
We call this procedure the ``X-fit'', and we repeat it for
each of the values of $am_y$.

We then proceed to the ``Y-fit'', illustrated in the right
panel of Fig.~\ref{fig:su2-4x3y-nnlo}.
Here we use either a linear or a quadratic fit to extrapolate
the short distance to the physical value
$Y_\TP = 2M_K^2 - M_\pi^2$.
We find, as seen in the figure, that the dependence is weak
and close to linear. Thus we use the linear fit for our central
value and the quadratic fit to estimate a systematic error.
In the figure, the red diamond shows the result
of linear extrapolation.

\section{Dependence on the Light Sea Quark Mass}
After the X- and Y-fits, the resulting intermediate value for $B_K$ 
still has the analytic dependence on the light sea-quark mass  $a m_l$
(entering through the $d_3$ term), as well as taste-conserving
discretization and truncation errors.
Here, we investigate the dependence on $am_l \propto L_\TP$,
which we have studied on the MILC coarse ensembles.
We also discuss the corresponding dependence of the result of
the SU(3) SChPT fit discussed in Ref.~\cite{ref:wlee:2009-1}.

\begin{figure}[t!]
\centering
\includegraphics[width=0.49\textwidth]{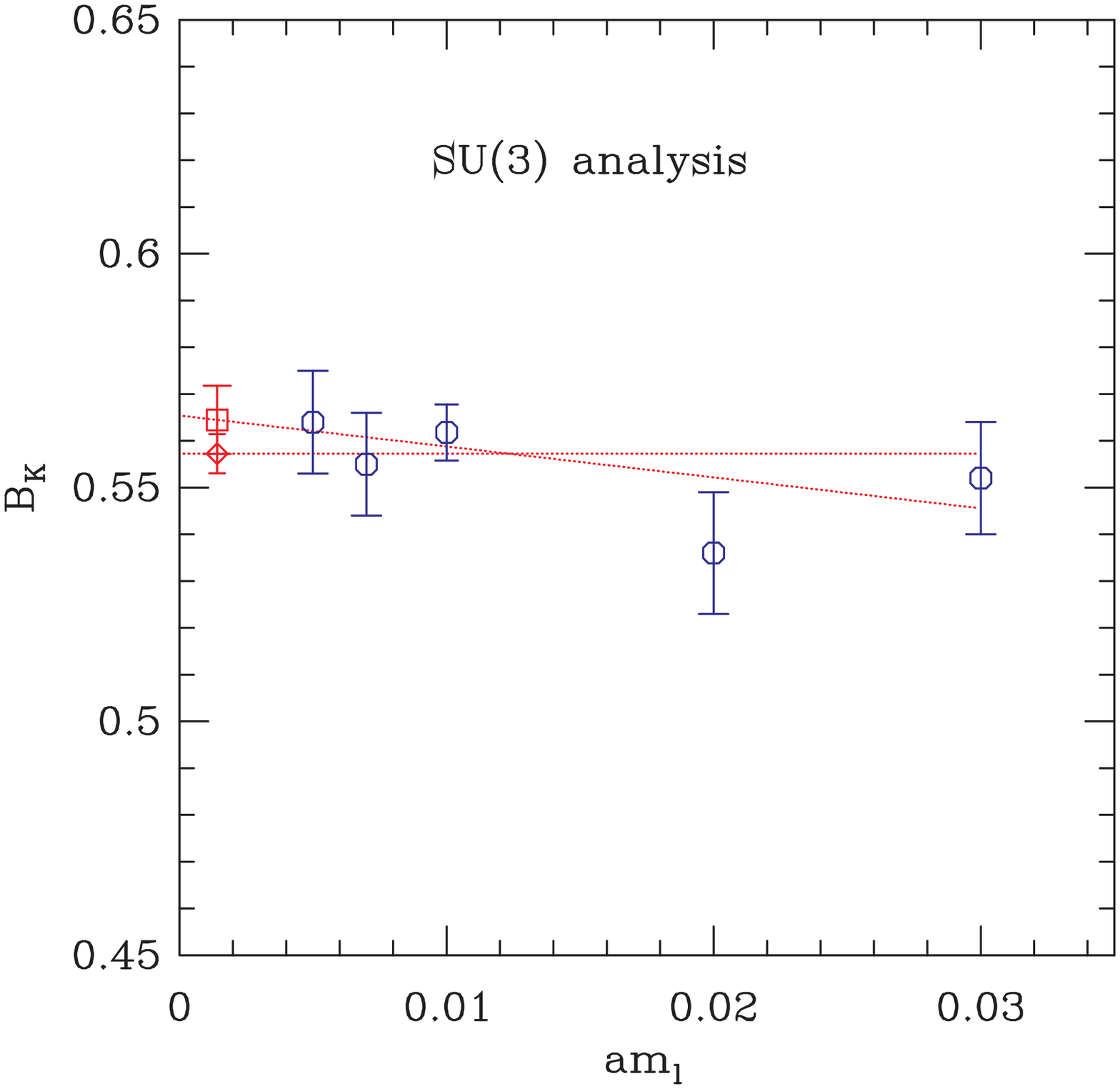}
\includegraphics[width=0.49\textwidth]{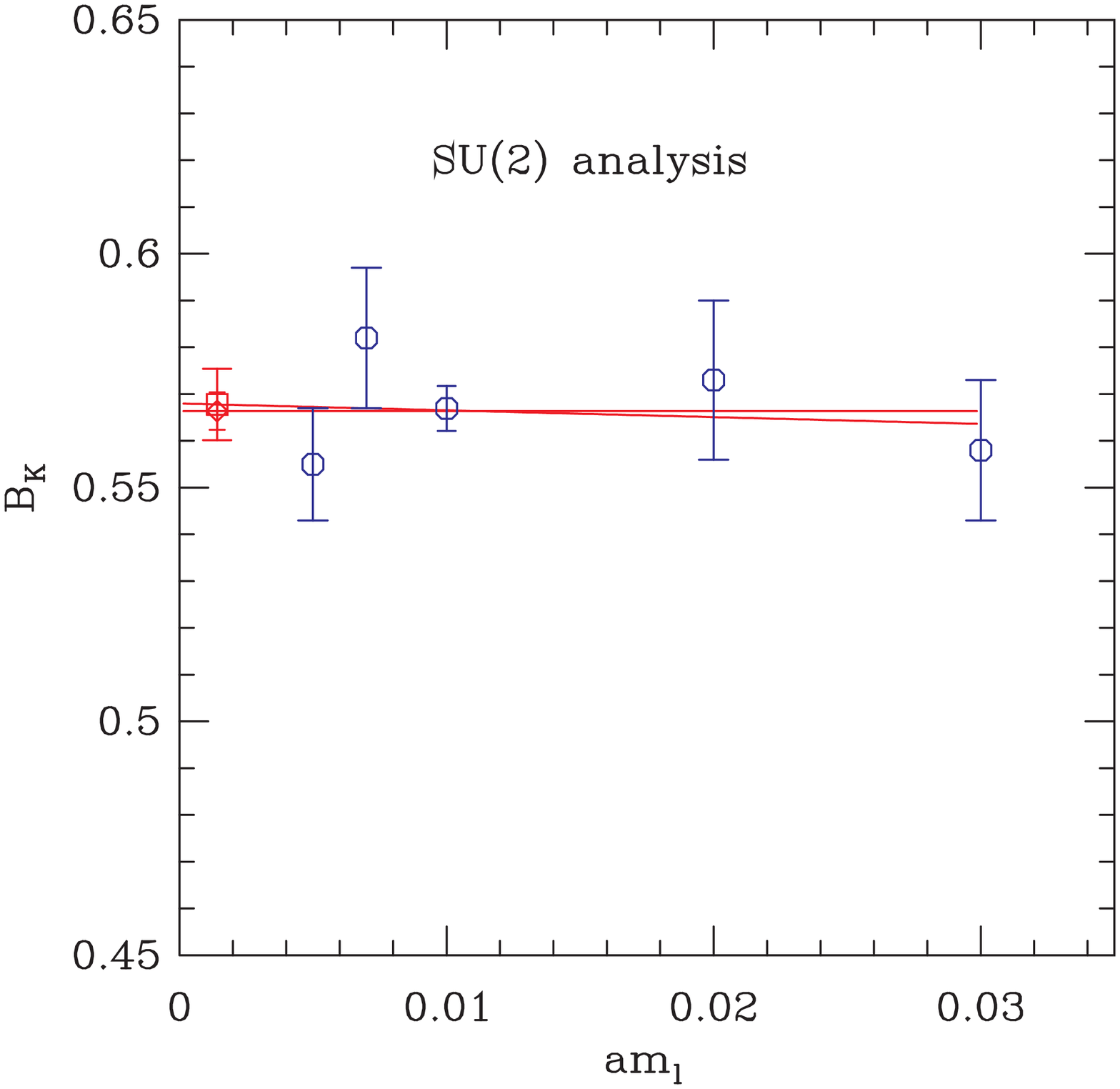}
\caption{ $B_K$ (after X- and Y-fitting) versus $am_l$ 
  for the SU(3) analysis (left) and for the
  SU(2) analysis (right). Data is from the MILC coarse
  lattices. The fit types are N-BT7 for SU(3) and 4X3Y-NNLO for
  SU(2). See text for details. }
\label{fig:bk:am_l}
\end{figure}
In Fig.~\ref{fig:bk:am_l}, we show the $am_\ell$ dependence
for both SU(3) and SU(2) fits.
We have fit to the expected linear dependence [resulting in the (red) squares],
and also, since the data show very little dependence on $a m_\ell$,
to a constant [(red) diamonds].

The weakness of the dependence on $a m_\ell$ is striking.
It has the important consequence that our use of only a single light
sea-quark mass on the fine and superfine lattices does not introduce
a large uncertainty. We use the difference between
the constant and linear fits as an estimate of the systematic error
arising from the uncertainty in the $a m_\ell$ dependence.
As the figure shows, this error is somewhat smaller for 
SU(2) fitting than for SU(3).

\section{Continuum Extrapolation}
\begin{figure}[t!]
\centering
    \includegraphics[width=0.49\textwidth]
                    {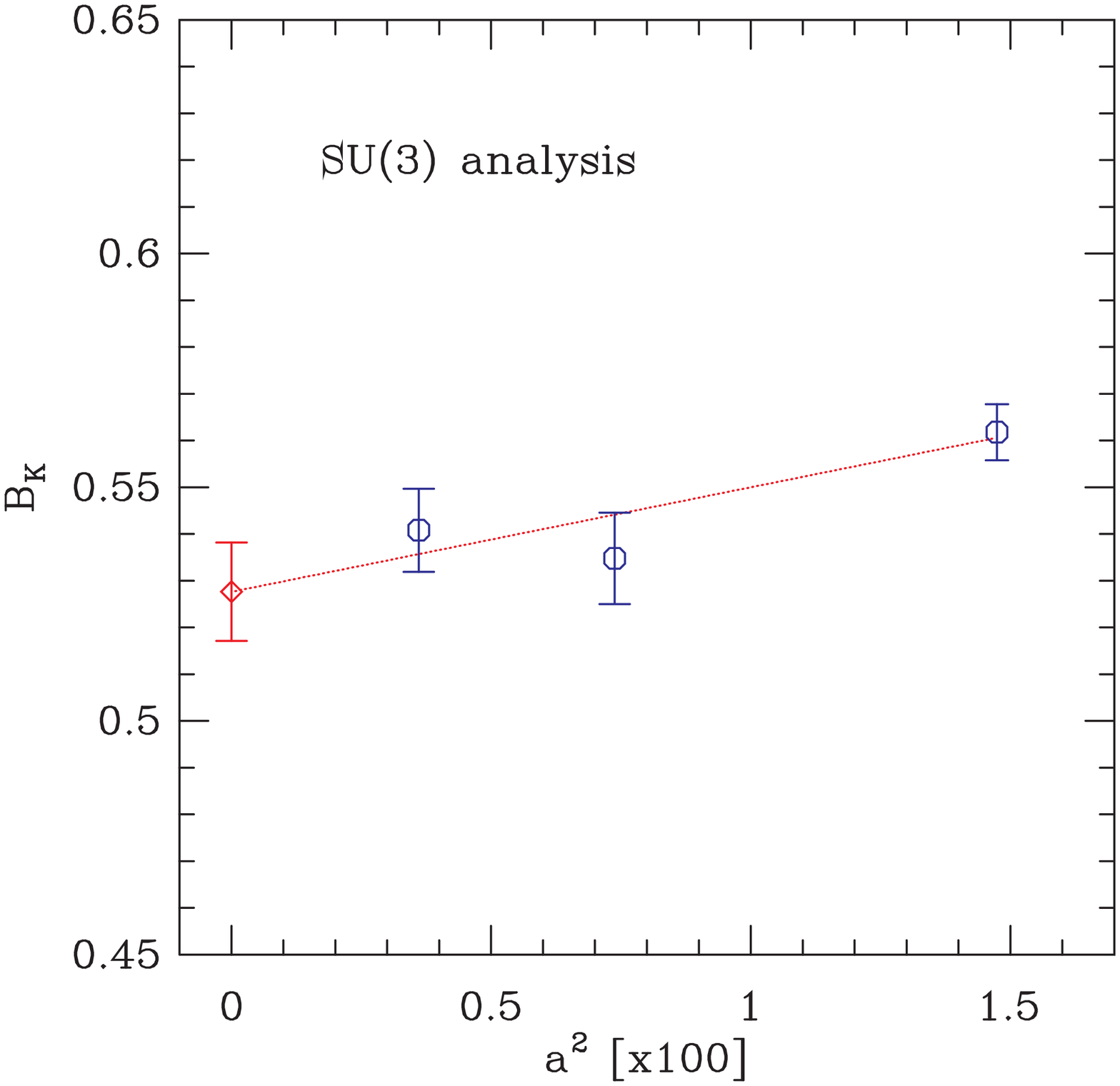}
    \includegraphics[width=0.49\textwidth]
                    {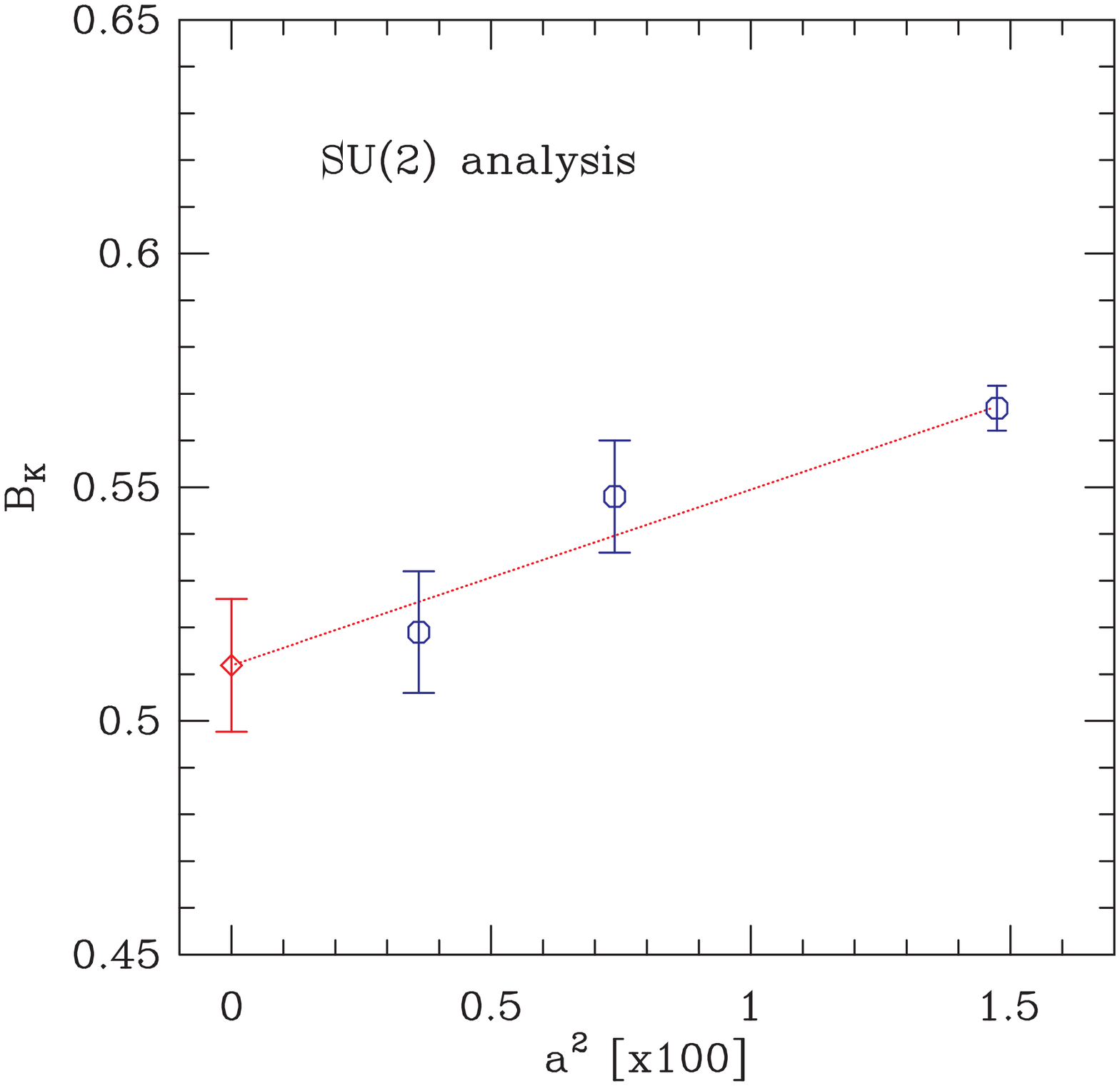}
\caption{One-loop matched $B_K$ plotted versus $a^2$ for 
the SU(3) analysis (left) and for the
  SU(2) analysis (right). The fit types are N-BT7 for SU(3) and
  4X3Y-NNLO for SU(2).}
\label{fig:bk:a^2}
\end{figure}

The dominant errors remaining at this stage are those due to taste-conserving
discretization and truncation errors.
These vary as $a^2\alpha^n$, where $n=0,1,\dots$
($n=0$ is allowed since we do not use Symanzik-improved operators), 
and as $\alpha^2$.
We cannot disentangle these effects using a fit to three lattice
spacings, so proceed as follows. We fit our data to a linear
function of $a^2$, and estimate the $O(\alpha^2)$ truncation error
separately (as described in a companion proceeding~\cite{ref:wlee:2009-4}).
More precisely, we fit to the data from the lattices with
$a=0.12, 0.09, 0.06\;$fm 
having $am_l/am_s$ fixed to $1/5$.
The results for both SU(3) and SU(2) analyses are shown in 
Fig.~\ref{fig:bk:a^2}.
We use the extrapolated values for our main result, and 
take the difference between them and the results on the
superfine lattices as an estimate of the systematic error
due to the continuum extrapolation.

The dependence on $a^2$ is noticeable for both analyses, with
$B_K$ increasing by about 6\% and 10\% between $a=0$ and $0.12\;$fm
in the SU(3) and SU(2) cases, respectively.
Assuming a form $B_K^{\rm cont}[1 + (a\Lambda)^2]$ this corresponds
to scales of $\Lambda\approx 400 $ and $500\;$MeV.
These are reasonable values, indicating that HYP-smearing has reduced
the discretization errors with staggered fermions down to
canonical size.

\section{Error Budget and Conclusion}
%
%
%
\begin{table}[tbhp]
\centering
\begin{tabular}{ l | l l }
\hline \hline
cause & error (\%) & description \\
\hline
statistics      & 2.8    & 4X3Y-NNLO fit (ensemble C3) \\
discretization  & 1.4    & diff. of (S1) and $a=0$ \\
fitting (1)     & 0.15   & X-fit: NLO vs. NNLO \\
fitting (2)     & 0.5    & Y-fit: linear vs. quadratic \\
fitting (3)     & 0.25   & constant vs linear $am_l$ dependence\\
finite volume   & 0.89   & $20^3$ (C3) versus $28^3$ (C3-2) \\
matching factor & 6.4    & $\Delta B_K^{(2)'}$ (S1) \\
$r_1$           & 0.09  & uncertainty in $r_1$ \\
\hline \hline
\end{tabular}
\caption{Preliminary error budget for $B_K$ obtained using 
 SU(2) SChPT fitting.
  \label{tab:su2-err-budget}}
\end{table}
In Table \ref{tab:su2-err-budget}, we summarize our present
best estimates of the uncertainties in $B_K$ arising from various sources.
The method by which we estimate these errors is outlined
in the ``description'', and has in most cases been explained above.
The statistical error is obtained from using a global jackknife
procedure.
The finite volume estimate
is obtained by comparing results on two volumes, as described
in one of the companion proceedings~\cite{ref:wlee:2009-3}.
This error is comparable to that estimated using NLO ChPT.
The error due to the use of one-loop  matching is estimated
in another of the companion proceedings~\cite{ref:wlee:2009-4}.
The error due to the uncertainty in the scale $r_1$ 
is estimated by varying the input values within the quoted
errors and repeating the analysis.

One error not included in this budget is that due to the
strange sea-quark mass differing slightly from its physical value.
Given the weak dependence on the light sea-quark mass, we expect
this to be a negligible effect, but plan to make a more
quantitative estimate in the future.

Combining these errors, our current, preliminary result for
$B_K$ using SU(2) SChPT fitting is
\begin{equation}
\begin{array}{l l c}
 B_K(\text{NDR}, \mu = 2 \text{ GeV}) & = 0.512 \pm 0.014 \pm 0.034 & 
\qquad [{\rm SU(2),\ {\bf PRELIMINARY}}]\,, \\
 \hat{B}_K = B_K(\text{RGI}) & = 0.701 \pm 0.019 \pm 0.047 & 
\qquad [{\rm SU(2),\ {\bf PRELIMINARY}}]\,,
\end{array}
\end{equation}
where the first error is statistical and the second systematic.
The total error is thus about 7\%, and is dominated by the
error in the matching factor. 
The total error is smaller than the 9\% preliminary error we find for
the SU(3) analysis~\cite{ref:wlee:2009-1}.
The statistical error in the latter analysis is smaller
(as can be seen in the earlier plots, and arises because
there are more data points in the fits)
but this is more than balanced by an increase in the
systematic errors.

Our result is consistent
with other $2+1$-flavor unquenched results obtained
using different fermion discretizations, 
all of which are summarized in Ref.~\cite{lubicz-2009-1}.
For example the calculation of Ref.~\cite{ALV} using
domain-wall valence fermions on the coarse and fine MILC
lattices finds 
$B_K({\rm NDR},\mu=2\;{\rm GeV}) = 0.527(6)(20)$.
The agreement between calculations using different
fermion discretizations provides a highly non-trivial
cross-check on the results.

During the next year, we aim to reduce our errors by
adding a fourth (``ultrafine'') lattice spacing ($a\approx 0.045\;$fm),
adding other sea-quark masses, and
by increasing the statistics.
We are also working on two-loop and non-perturbative calculations of
the matching factors.

\section{Acknowledgments}
C.~Jung is supported by the US DOE under contract DE-AC02-98CH10886.
The research of W.~Lee is supported by the Creative Research
Initiatives Program (3348-20090015) of the NRF grant funded by the
Korean government (MEST). 
The work of S.~Sharpe is supported in part by the US DOE grant
no.~DE-FG02-96ER40956. Computations were carried out
in part on facilities of the USQCD Collaboration,
which are funded by the Office of Science of the
U.S. Department of Energy.

\end{document}